\documentclass[12pt]{article}
\let\section=\subsection     \let\subsection=\subsubsection                
\newcommand{\dsl}[1]{#1 \hspace{-0.15cm}\slash }
\newcommand{\imag}[0]{i}
\newcommand{\fourint}[1]{\int\!\frac{d^4 #1}{(2\pi)^4}}

\usepackage{ae,graphicx,epsfig}

\begin{document}
\begin{center}
   {\large \bf NUCLEONS IN THE COVARIANT QUARK--DIQUARK MODEL}\\[5mm]
   M.~OETTEL\footnote{\em Feodor-von-Lynen fellow, present address:
 CSSM, 10 Pulteney St, Adelaide, SA 5005, Australia}  \\[5mm]
   {\small \it  Institute for Theoretical Physics, T\"ubingen University \\
   Auf der Morgenstelle 14, D-72076 T\"ubingen, Germany \\[8mm] }
\end{center}

\begin{abstract}\noindent
We introduce diquarks as separable correlations in the two--quark
Green's function to facilitate the description of baryons as relativistic
three--quark bound states. These states then emerge as solutions of
Bethe--Salpeter equations for quarks and diquarks that interact via
quark exchange. Approximating quark and diquark propagators by the 
corresponding free ones, we calculate nucleon static properties and 
form factors. For the description of production 
processes off the nucleon, we consider various dressing functions 
for the propagators to remove unphysical thresholds. Results
for kaon
photoproduction, $\gamma p\to K\Lambda$, and associated strangeness
production, $pp\to pK\Lambda$, allow us to draw conclusions
on the permissibility of different dressing functions.
\end{abstract}

\section{The Covariant Quark--Diquark Model}

In the last few years enormous experimental progress in studying hadronic
properties in the few GeV regime has been made
that encourages the study of relativistic, explicitly
covariant models of baryons. We adopt a Green's function approach
and study the 3-quark correlation function whose poles signal
the appearance of bound states, the baryons. 
To this end, the propagation of a single quark and of two quarks
need to be investigated.
For single quarks,
it has been confirmed
by Dyson--Schwinger studies \cite{Roberts:2000aa,Alkofer:2000wg} 
that they acquire a dynamically generated
constituent mass by gluon dressing. 
Additionally these studies indicate that the quark propagator does not possess
poles for real values of its momentum squared and is thus confined. 
Therefore we parameterize
the single quark propagator (in Euclidean space) by
\begin{eqnarray}
 S^{(k)} (p) &=& \frac{\imag \dsl{p}-m_q }{p^2+m_q^2}  \;
   f_{k}\left(\frac{p^2}{m^2_q}\right) \; ,
\label{sk}
\end{eqnarray}
Quarks have a constituent mass $m_q \approx 400$ MeV. We remark that
for setting the dressing function $f_0=1$, eq.~(\ref{sk}) describes a free quark.
Absence of real poles can be achieved by the choices \cite{Ahlig:2000qu}
\begin{eqnarray} \label{fdef}
 &&f_1 (x) = \frac{1}{2}\left\{
  \frac{x+1}{x+1-i/d} + \frac{x+1}{x+1+i/d} \right\}\;, \\
 && f_2 (x) =  1- \exp\left[ -d\left(1+x \right) \right]\; , \quad
  f_3 (x) =  \tanh\left[ d\left(1+x\right)
         \left(1+x^\ast\right) \right] \; . \nonumber
\end{eqnarray} 
For the first choice, the free mass pole is shifted to a pair
of complex conjugate poles thereby mimicking creation and decay processes
of quarks that cancel each other in physical processes. The second choice
models the propagator as an entire analytical function, at the expense
of an essential singularity at timelike infinity, $p^2 \to -\infty$.
Applications of this form may be found in ref.~\cite{Roberts:2000aa}.
The third form prescribes a pole free form with an asymptotic free particle
behavior in all directions of the complex momentum plane. This can 
be achieved only by adopting a non--analytic dressing function.

Equipped with models for the single quark propagator we investigate the
relativistic 3--quark problem. We neglect any three--particle
irreducible interaction graphs between the quarks, which defines the well--known
Faddeev problem.  For the 2--quark correlations,
we observe that any two quarks 
within baryons have to be in a color 
$\bar {\bf 3}^c$ representation and that the gluon exchange between quarks
in the representation $\bar {\bf 3}^c$ is attractive. Furthermore lattice results
indicate strong diquark correlations in the scalar channel \cite{Hess:1998sd}.
This motivates a separable {\em ansatz} for the quark--quark
$t$--matrix of the form
\begin{equation}
t_{\alpha\gamma , \beta\delta}^{\hbox{\tiny sep}}(p,q,P) \, =
\chi_{\gamma\alpha}(p) \,D(P)\,\bar \chi_{\beta\delta}(q) \; +\;
\chi_{\gamma\alpha}^\mu(p) \,D^{\mu\nu}(P)
    \bar  \chi_{\beta\delta}^\nu (q)  \, .
\label{Gsep}
\end{equation}
Here, $P$ is the total momentum of the incoming and the outgoing quark-quark
pair, $p$ and $q$ are the relative momenta between the quarks.
$\chi_{\alpha\beta}(p)$ and $\chi_{\alpha\beta}^\mu(p)$ are
vertex functions of quarks with a scalar and an axialvector
diquark, respectively. 
We parameterize the finite size of the diquark vertices by a dipole form.
We take the associated width parameter, that directly
influences the proton electric radius, to be of the order
300--400 MeV.
The inclusion of axialvector diquarks is the minimal
requirement to describe decuplet baryons and, as it turns out later, is
crucial for describing the nucleon electromagnetic form factors correctly.
For simplicity, the diquark propagators $D^{[\mu\nu]}$ are taken to
be free propagators of a spin--0 [spin--1] particle multiplied by
the dressing functions defined in eq.~(\ref{fdef}). 
 
\begin{figure}
\centerline{\epsfig{file=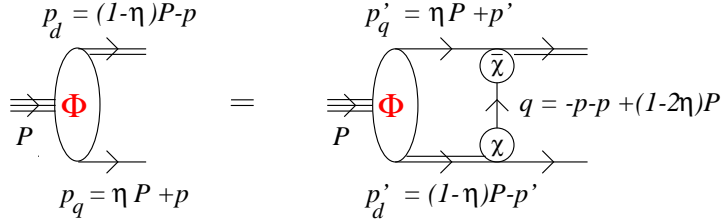, width=10cm}}
\caption{The baryon Bethe-Salpeter equation. The momentum partitioning
parameter $\eta$ distributes the relative momentum $p$ ($p'$)  over
quark and diquark.}
\label{bse}
\end{figure}

Having imposed the separable {\em ansatz} (\ref{Gsep}) the Faddeev equations reduce to a coupled system
of Bethe--Salpeter equations describing baryons as bound states
of quarks and diquarks which interact by quark exchange. This interaction is by virtue
of the color degree of freedom attractive and restores the Pauli principle. 
For the nucleon these equations reads
\begin{eqnarray}
  \label{eq: BS_1}
  \fourint{p'} K(p,p',P)
  \pmatrix{\Psi^5 \cr \Psi^{\mu'}}(p',P) &=& 0
\end{eqnarray}
The interaction part of the kernel $K$,
\begin{eqnarray}
 K (p,p',P) &=& (2\pi)^4\: \delta(p-p')\: S^{-1}(p_q)
  \pmatrix{ D^{-1} & 0 \cr 0 & (D^{-1})^{\mu'\mu}} (p_d) + \nonumber \\
 & & \frac{1}{2}
  \pmatrix{\chi\: S^T(q)\:\bar\chi & -\sqrt{3} \chi^{\mu'}\: S^T(q)\:\bar\chi \cr
    -\sqrt{3}\chi\: S^T(q)\:\bar\chi^{\mu} &  -\chi^{\mu'}\: S^T(q)\:\bar\chi^\mu} \; ,
 \label{Gdef}
\end{eqnarray}
is given by the quark exchange (for the definition of the involved
momenta see Fig.~\ref{bse}). We have solved these equations 
without further reduction and thus obtained
covariant spinorial wave functions  $\Psi^{[\mu]}$. 

In a study employing free quark and diquark propagators \cite{Oettel:2000jj}
we calculated the nucleon electromagnetic form factors. Gauge invariance
and correct charge and norm were guaranteed by coupling the photon to
all possible places in the kernel of the Bethe--Salpeter equation
given in eq.~(\ref{Gdef}). The results for the electric form factors 
(up to momentum transfers of 2.5 GeV$^2$)
are in good agreement with the experimental data, nevertheless
it turned out  to be impossible to obtain a simultaneous correct description
of the nucleon magnetic moments and the mass of the $\Delta$ isobar.
Due to the free particle thresholds $m_q>411$ MeV had to be chosen
to obtain a bound $\Delta$ and these constituent quark masses
yielded proton magnetic moments $\mu_p \approx 1.9$. For lower
masses $m_q=360$ MeV we found $\mu_p \approx 2.5$, thus illustrating the
necessity to avoid the free--particle poles for the quarks.
We found that
20--25~\% axialvector correlations (measured by the ratio of the 
norm contributions stemming from $\Psi$ and $\Psi^\mu$)
are needed to describe the ratio
of proton electric to magnetic form factor in accordance with experiment.

\section{Production Processes and Quark Confinement}

\begin{figure}
\centerline{
    \epsfig{file=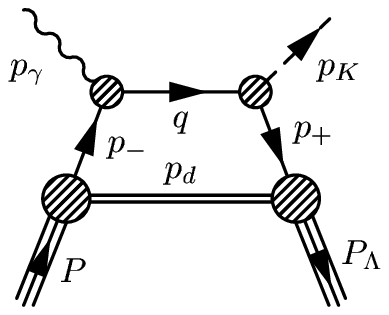,width=5cm,height=4cm}
\hspace{1cm}
    \epsfig{file=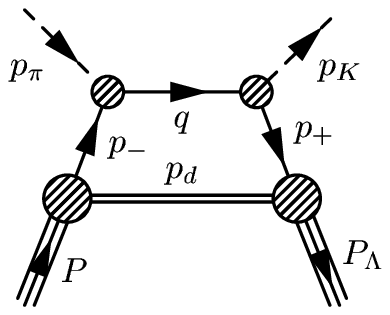,width=5cm,height=4cm}
}
    \caption{ {\it Left panel:} Main contribution to 
    $p\gamma\to K\Lambda$, {\it right panel:} Handbag
    diagram contributing to the reaction $pp \rightarrow pK\Lambda$ as
    a subprocess. The incoming pion couples to the `spectating' proton.}
    \label{kaon_hand_bag}
\end{figure}

In ref.~\cite{Ahlig:2000qu} we studied the processes $\gamma p\to K\Lambda$
and $pp\to pK\Lambda$ within the covariant diquark model in impulse
approximation. These processes have been investigated experimentally 
\cite{Tran:1998qw,COSY-TOF}. Assuming for the 
moment just scalar correlations to be present in $p$ and $\Lambda$,
the impulse approximation for both processes is restricted to diquark
spectator graphs of which the handbag--type ones 
are shown in Fig.~\ref{kaon_hand_bag}. A kinematical analysis
shows that the momentum $q$ of the intermediate quark is far in the 
timelike domain, thus, using  free quarks and diquarks, amplitudes corresponding 
to the handbag diagrams would show unphysical thresholds. For this reason
the dressing functions of eq.~(\ref{fdef}) have been introduced and 
the quality of their parameterization of the timelike domain
can be assessed by studying the above mentioned production processes. 

We solved the Bethe--Salpeter equations for nucleon, $\Lambda$ and the 
other octet baryons using the dressed propagators. For the
non--analytic form ($f_3$) we found a 
violation of relativistic translation invariance.
We calculated the electromagnetic form factors to further constrain
model parameters. The formalism shortly described in the previous section
had to be adapted to describe gauge--invariantly the photon
couplings to quarks and diquarks. For the case of 
non--analytic propagators we found that the well--known Ball--Chiu
construction (see {\em e.g.} refs.~\cite{Roberts:2000aa,Alkofer:2000wg}) 
had to be modified in a way that depends on the frame
that is used to calculate the form factors. Therefore already at this level 
it is safe to conclude that non--analytic propagators can be excluded
in the search for an effective confinement parameterization due to the
problems with relativistic invariance.   

\begin{figure}
 \begin{center}
  \epsfig{file=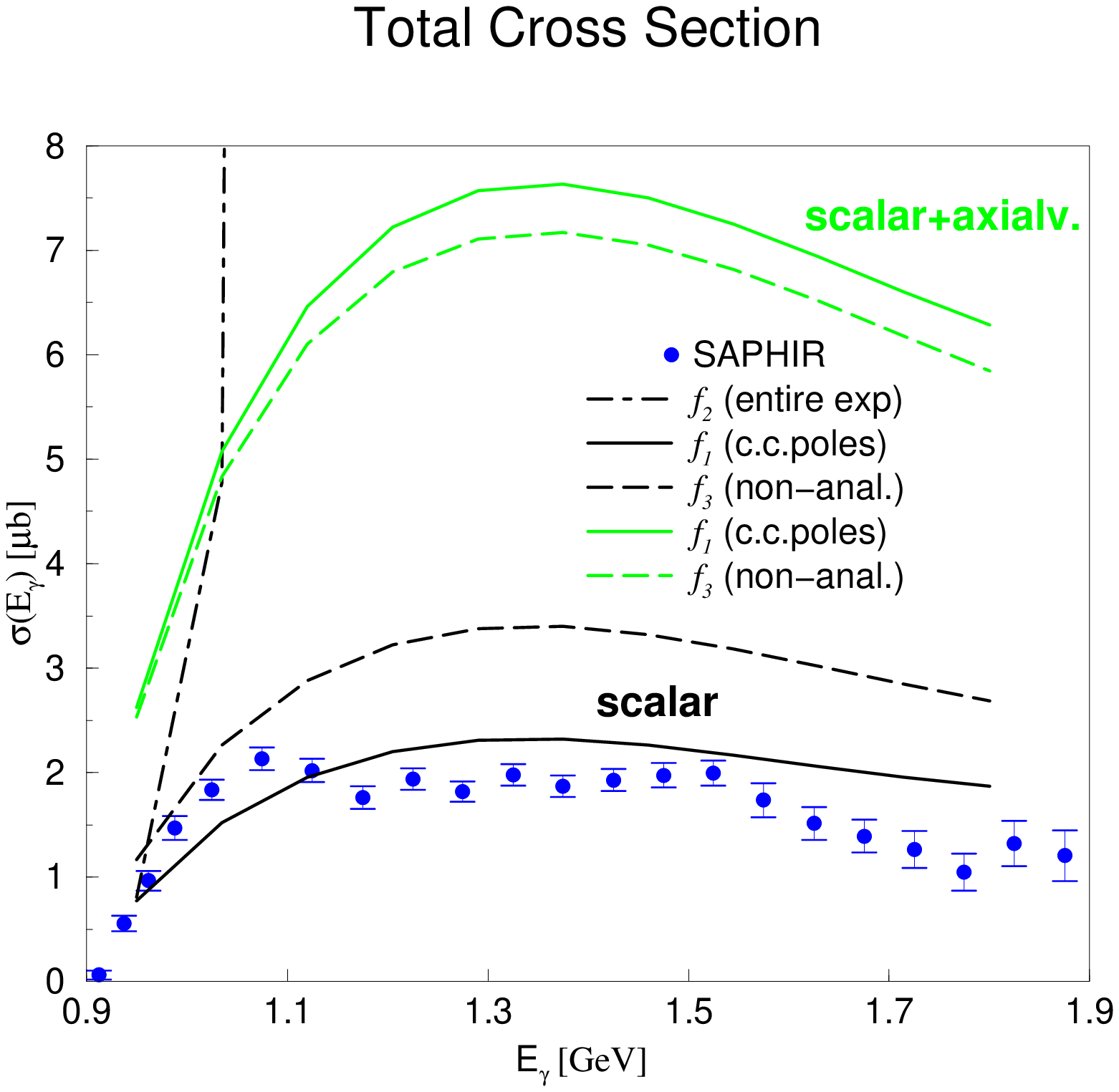,width=6.3cm} 
  \epsfig{file=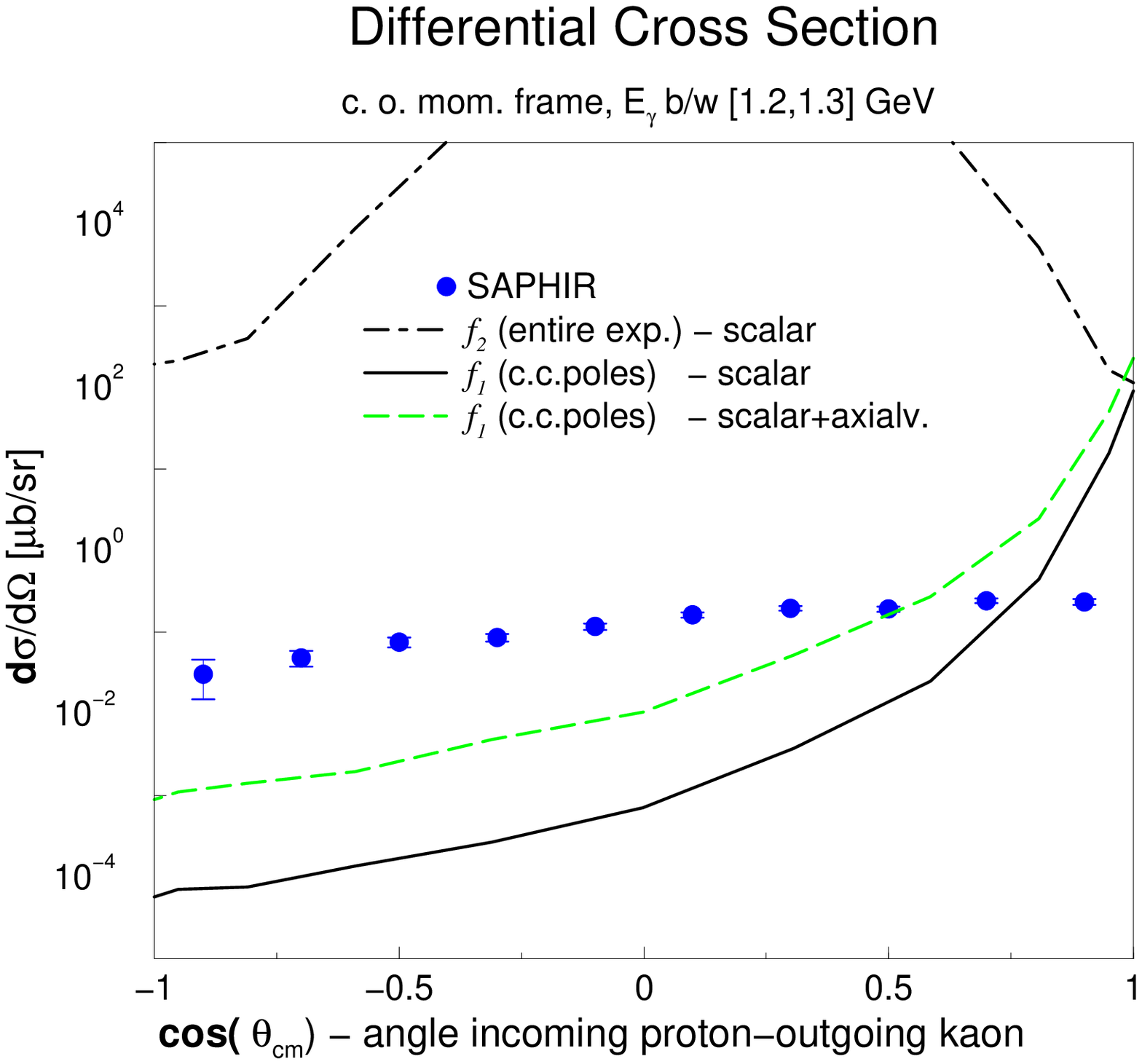,width=6.3cm}
 \end{center}
 \vspace{-0.5cm}
 \caption{ The process $\gamma p \to K\Lambda$. {\it Left panel:}
 Total cross section {\it vs.} the energy of the incoming photon.
 The black curves show results for data sets with scalar diquark correlations
 only but with different propagator dressing functions. The light curves
 show results when axialvector correlations are included using the dressing
 functions $f_1$ and $f_3$. 
 {\it Right panel:} Differential cross section in the center--of--momentum
 frame.}
 \label{photo_pics}
\end{figure}

\begin{figure}[t]
 \begin{center}
  \epsfig{file=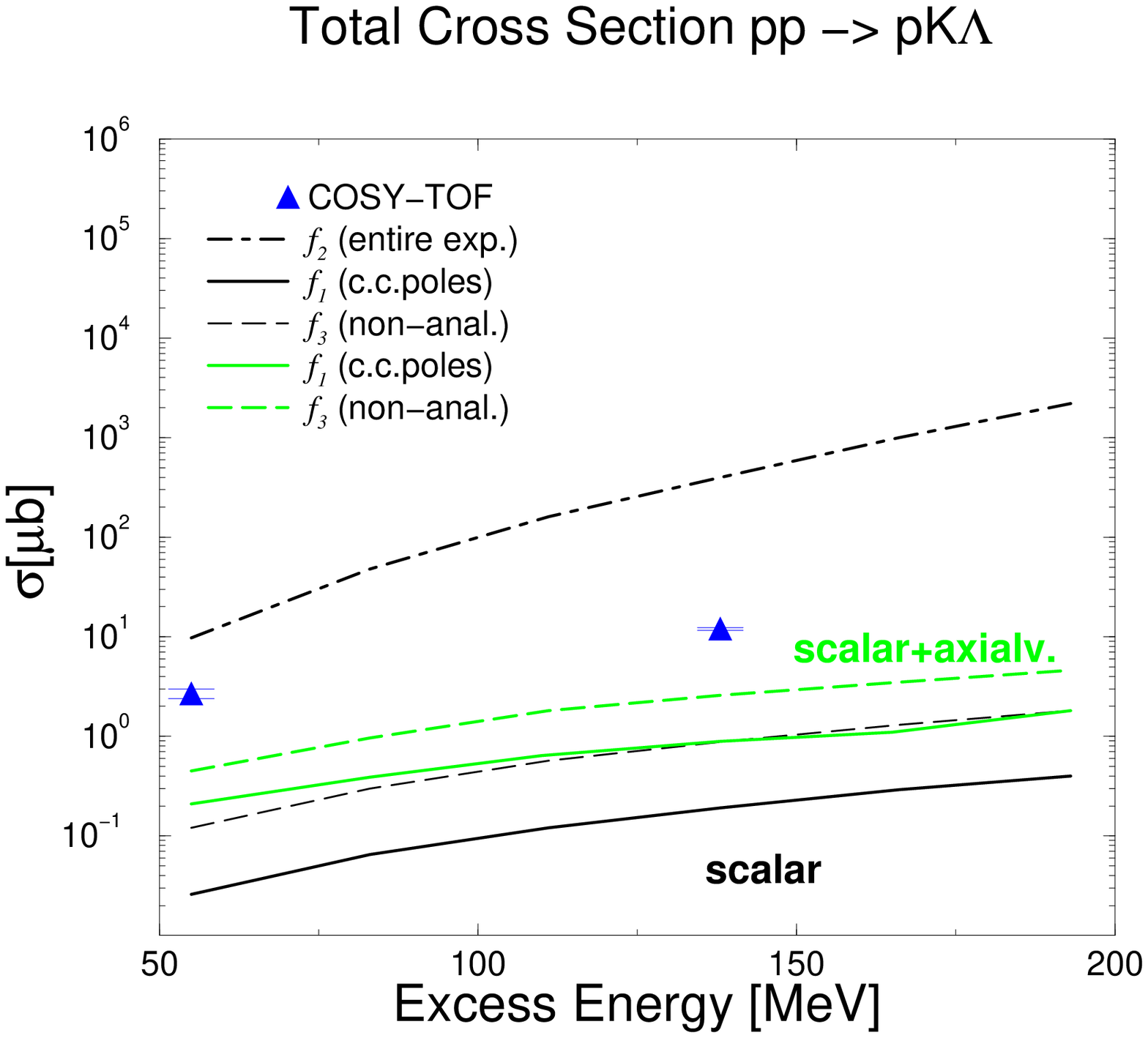,width=6.3cm} 
  \epsfig{file=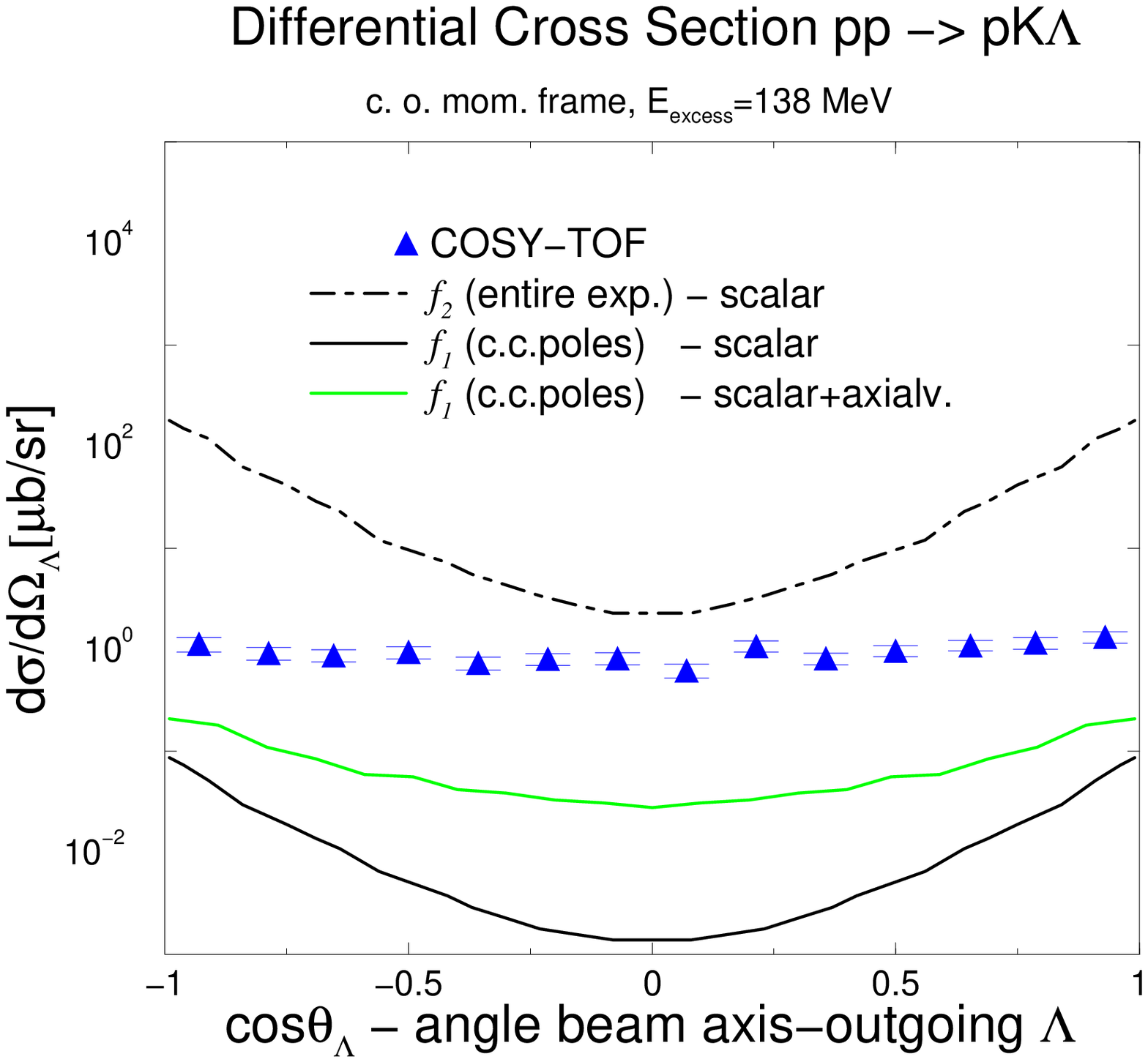,width=6.3cm}
 \end{center}
 \vspace{-0.5cm}
 \caption{ The process $p p \to p K\Lambda$. {\it Left panel:}
 Total cross section {\it vs.} the excess energy.
 {\it Right panel:} Differential cross section in the center--of--momentum
 frame at an excess energy of 138 MeV.} 
 \label{ass_pics}
\end{figure}

Dressing the quark--photon vertices leads to an enhancement of 
$\mu_p$, being close to the experimental value for all choices of the
dressing function.
The inclusion of 20 \% axialvector correlations in the nucleon yields
considerable improvements on the ratios $\mu_p/\mu_n$, $G_E/G_M$
and the values of the $\Sigma$ and $\Xi$ hyperon masses. Therefore
we have extended the study of the production processes to impulse
approximation diagrams with the quark being spectator. These arise 
for a non--zero axialvector admixture in the baryon wave function. The results for kaon photoproduction
are shown in Fig.~\ref{photo_pics}. Here the most striking observation
is the drastic increase of the cross section above  threshold for
the choice of the entire analytical function $f_2$.
The latter is enhanced in the timelike region and thus the handbag diagram
of figure \ref{kaon_hand_bag} dominates, yielding completely unphysical
results. For the other two choices, $f_1$ and $f_3$, the handbag diagram
yields negligible cross sections compared to a $t$--channel $K$ exchange
diagram. 
The overestimation of the cross section for the data sets using scalar and
axialvector correlations is due to an oversimplified treatment
of the kaon--quark vertex \cite{Ahlig:2000qu}. 
The results for the process $pp \to pK\Lambda$,
given in Fig.~\ref{ass_pics}, show similar, unphysical enhancement
of the cross section for the choice $f_2$ for the dressing function.
For the other choices, total cross sections are too small and 
differential cross sections show a dip which is not seen in experiment. This
indicates shortcomings of the impulse approximation, however, as for the form 
factors, the inclusion of axialvector correlations in the baryons yields
considerable improvement. 

In summary, a picture of the nucleon including
scalar and axialvector diquark correlations together with an effective parameterization
of confinement in form of dressing functions with complex conjugate poles seems
to be worth further study.

\noindent
{\bf Acknowledgements:} The author wants to express his gratitude to 
the organizers for the invitation
and for staging this pleasant workshop. 
He thanks S. Schmidt for support and his co--workers
S.~Ahlig, R.~Alkofer, C.~Fischer, H.~Reinhardt, L.~v.~Smekal and H.~Weigel.
This work was supported by
the DFG (We 1254/4-2 and Schm 1342/3-1)
and COSY (4137660).


\end{document}